\documentclass[aps,twocolumn, superscriptaddress, svgnames]{revtex4-2}
\usepackage[T1]{fontenc} % Use modern font encodings
\usepackage{tikz,tikzscale, relsize, pgfplots, xcolor, amssymb, amsmath}
\usetikzlibrary{backgrounds, arrows, calc, decorations, positioning, decorations.pathmorphing,decorations.pathreplacing, 
	decorations.markings, fixedpointarithmetic, fit, patterns, decorations.text, backgrounds,matrix,plotmarks, spy, shadings}
\usepgfplotslibrary{fillbetween}
\pgfplotsset{compat=1.17}

\usepackage{graphicx}
\newcommand{\RomanNumeralCaps}[1]
    {\MakeUppercase{\romannumeral #1}}

\usepackage{capt-of}

\begin{document}
\title{Geometry Dependent Localization of Surface Plasmons on Random Gold Nanoparticle Assemblies}

\author{Mohammed Fayis Kalady}
\email{m.f.kalady@ifw-dresden.de}
\affiliation{Leibniz Institute for Solid State and Materials Research (IFW) Dresden, Helmholtzstraße 20, 01069 Dresden, Germany}
\affiliation{Indian Institute of Technology (IIT) Delhi, Hauz Khas, New Delhi, Delhi 110016, India}
\author{Johannes Schultz}

\affiliation{Leibniz Institute for Solid State and Materials Research (IFW) Dresden, Helmholtzstraße 20, 01069 Dresden, Germany}
\author{Kristina Weinel}

\affiliation{Leibniz Institute for Solid State and Materials Research (IFW) Dresden, Helmholtzstraße 20, 01069 Dresden, Germany}
\affiliation{Federal Institute of Materials Research and Testing (BAM), Unter den Eichen 87, 12205 Berlin, Germany}
\author{Daniel Wolf}

\affiliation{Leibniz Institute for Solid State and Materials Research (IFW) Dresden, Helmholtzstraße 20, 01069 Dresden, Germany}
\author{Axel Lubk}
\email{a.lubk@ifw-dresden.de}
\affiliation{Leibniz Institute for Solid State and Materials Research (IFW) Dresden, Helmholtzstraße 20, 01069 Dresden, Germany}
\affiliation{Institute of Solid State and Materials Physics, TU Dresden, Haeckelstraße 3, 01069 Dresden, Germany}
%%%%%%%%%%%%%%%%%%%%%%%%%%%%%%%%%%%%%%%%%%%%%%%%%%%%%%%%%%%%%%%%%%%%%

\begin{abstract}

Assemblies of plasmonic nanoparticles (NPs) support hybridized modes of localized surface plasmons (LSPs), which delocalize in geometrically well-ordered arrangements. Here, the hybridization behavior of LSPs in geometrically completely disordered arrangements of Au NPs fabricated by an e-beam synthesis method is studied. Employing electron energy loss spectroscopy in a scanning transmission electron microscope in combination with numerical simulations, the disorder-driven spatial and spectral localization of the coupled LSP modes that depend on the NP thickness is revealed. Below 0.4\,nm sample thickness (flat NPs), localization increases towards higher hybridized LSP mode energies. In comparison, above 10\,nm thickness, a decrease of localization (an increase of delocalization) with higher mode energies is observed. In the intermediate thickness regime, a transition of the energy dependence of the localization between the two limiting cases, exhibiting a transition mode energy with minimal localization, is observed. This behavior is mainly driven by the energy and thickness dependence of the polarizability of the individual NPs.\\

\vspace{0.02in}

\noindent \textit{Keywords:\,}{Plasmonics, Electron-Energy Loss Spectroscopy, Discrete Dipole Approximation, Nanoparticles}
\end{abstract}

\maketitle
\section{Introduction}

Strongly disordered micro- and nanostructures of (noble) metals, such as random assemblies of gold nanoparticles (NPs) or thin random gold networks, are currently used for various applications such as electromechanical gas and vapor sensors \cite{Schlicke2017, Kim2019-sl} or flexible electrodes, e.g., for neural implants \cite{Minev2015} or monitoring of blood vessels \cite{Liu2017}. The latter exploits the presence of strongly localized surface plasmon (LSP) resonances in a particular frequency band depending on the composition and geometry of the nanostructure. Localization contrasts with the formation of delocalized modes in periodic structures (such as lattice resonances and quasicontinuous plasmon bands), which exhibit high quality factors, strong dispersion, and ballistic transport, among other characteristics \cite{Mayer2019, Kravets2018, Zou2006, Markel2007}. Both phenomena originate from the coupling and hybridization of LSP resonances via electromagnetic interaction \cite{Kravets2018}. Localization of surface plasmons in strongly coupling random networks and  NP assemblies has been intensely studied by various groups \cite{Schultz2024Anderson,Xomalis2021, Raguindin2023, Fregoni2021}, establishing the fundamental role of the disorder in the observed spatial localization and spectral distribution.  Recently, we have demonstrated the localization of surface plasmons in ultrathin random gold networks, which increased significantly in an energy interval from 0.3 to 1.6\,eV, and the absence of hybridized modes beyond \cite{Schultz2024Anderson}. In order to model the intriguing localization behavior, we described the random network as a random assembly of strongly oblate Au NPs (i.e., the inverted system in the sense that holes of the network correspond to NPs) by exploiting Babinet's principle \cite{Horak2019, Zentgraf2007, Falcone2004}, which was then amenable to numerical solutions. Notwithstanding obtaining good agreement between experiment and simulation, the question of how the thickness and other geometric parameters affected the localization behavior remained open.

\begin{figure}[!ht]
    \includegraphics{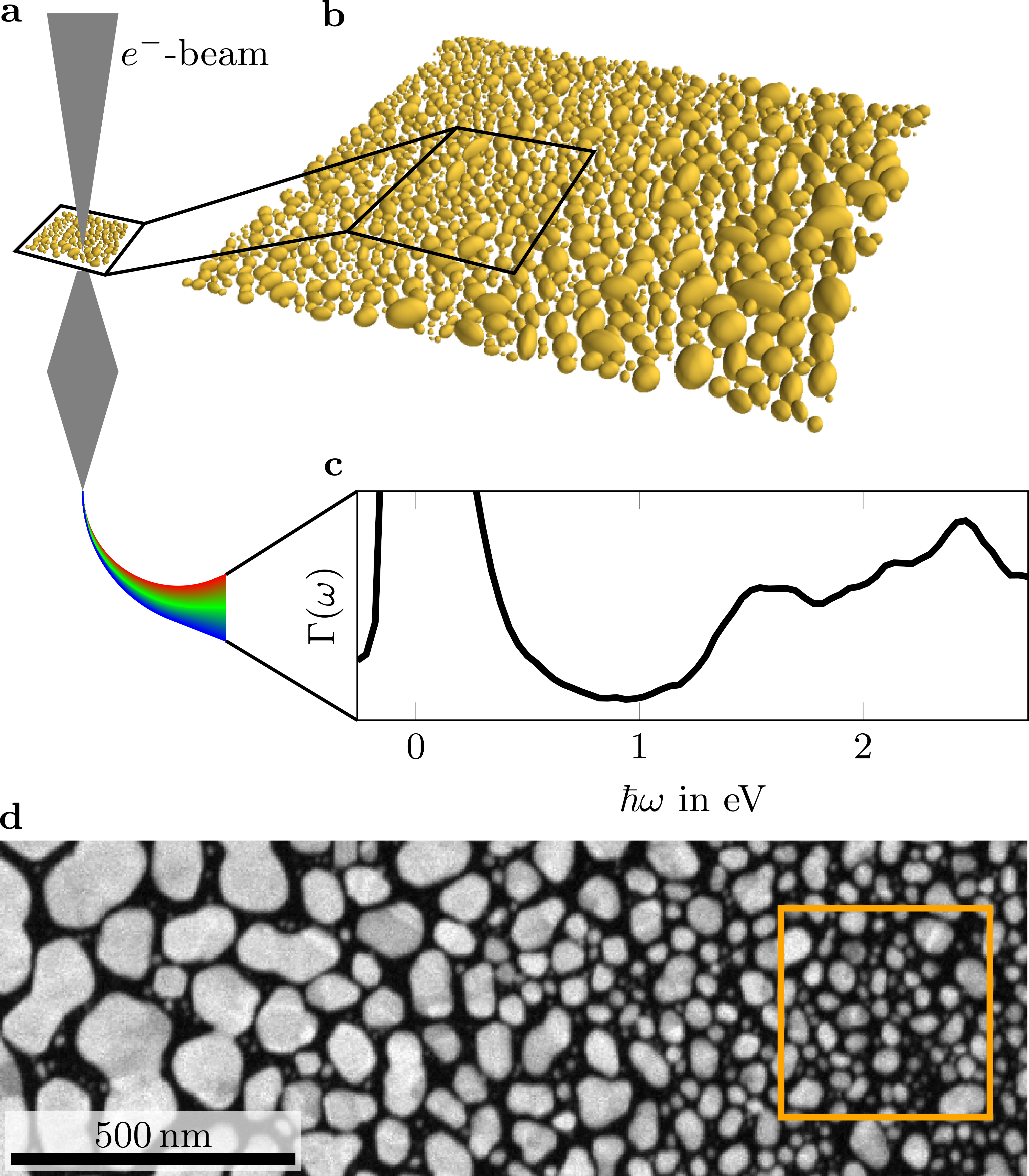}
\caption{ Plasmon mapping using STEM-EELS of randomly distributed gold NPs. (a) Scheme of a focused electron beam (grey) that is scanned over the sample (b), where the gold NPs are simplified as ellipsoids. (c) An EEL spectrum is collected at each scanning point within the black box region in (b). (d) HAADF STEM image of gold NP assembly depicted in (b) containing the orange indicated region that has been mapped by STEM-EELS yielding the average spectrum shown in (c).}
\label{fig:exp}
\end{figure}

In this work, we study the direct system, i.e., a random assembly of Au NPs fabricated by a recently developed synthesis route utilizing the electron beam within a scanning electron microscope (SEM) \cite{Weinel2024e}. This fabrication method has the advantage of synthesizing highly disordered assemblies of NPs of a large size distribution ranging from 2 to 100\,nm. We probe the spectral properties as well as spatial localization by spatially resolved electron energy loss spectroscopy (EELS) in the scanning transmission electron microscope (STEM) (see Fig.\,\ref{fig:exp} for a scheme). Numerical simulations have been performed using a self-consistent dipole model.  We find a localization behavior that decreases toward higher energies in contrast to the ultrathin networks, where it increased. The numerical simulations based on a discrete dipole approximation show that this qualitatively different behavior can be traced back to the difference in average thickness. 

%%%%%%%%%%%%%%%%%%%%%%%%%%%%%%%%%%%%%%%%%%%%%%%%%%%%%%%%%%%%%%%%%%%%%%%%%%%%%%%%%%%%%%%%%%%%%%%%%%%%%%%%%%%%%%%%%%%%
\section{Experiment}
A large number of ellipsoidal gold NPs with a wide range in size and, hence, thickness were fabricated using a synthesis approach described by Weinel et al. \cite{Weinel2024e}. This approach involves the partial sublimation of a gold precursor of approximately one micrometer size induced by the deposited energy of an intense electron beam within the SEM. The vaporized gold redeposits on an amorphous silicon oxide substrate as gold NPs with decreasing size at increasing distances to the precursor's position (see Fig.\,\ref{fig:exp} for a representative example). In order to study the dependency of the plasmon localization on the NPs' thickness, STEM-EELS datasets were recorded at spatial subsets of the Au NP distribution with particle sizes (see indicated region in Fig.\,\ref{fig:exp}d) similar to the hole sizes in the networks of our previous measurements report in \cite{Schultz2024Anderson}.\\
The STEM-EELS datasets were recorded using a FEI Titan$^3$ transmission electron microscope (TEM). During the scanning of a focused beam of electrons with 80\,keV kinetic energy, a High Angle Annular Dark Field (HAADF) image was acquired. At the same time, EEL spectra were recorded at different probe positions on the sample. Here, the signal corresponds to the probability $\Gamma$ of the beam electrons at probe position $(x,y)$ that suffer from a specific energy loss $\omega$, i.e., to excite a plasmon mode with a specific energy. This signal is called loss probability and corresponds to a projection of the plasmons' electric field component parallel to the electron beam: $\Gamma(x,y,\omega)\propto\int_{-\infty}^{\infty}\Re \left\lbrace e^{-\text{i}\omega z / v_z}\tilde{E}_z(\omega, x, y, z)\right\rbrace \text{d}z$.
The combination of EELS (spectral resolution) and STEM (spatial resolution) allows to study: \RomanNumeralCaps{1}) the spectral position and width of different plasmon modes in dependence of the probe position, \textit{i.e.}, $\Gamma(\omega)$ at fixed $(x,y)$ see Fig.\,\ref{fig:exp} (c) and \RomanNumeralCaps{2}) the spatial localization of plasmon modes from so-called loss probability maps ($\Gamma(x,y)$ at fixed $\omega$, see Fig.\,\ref{fig:exp_data} (a)).

%%%%%%%%%%%%%%%%%%%%%%%%%%%%%%%%%%%%%%%%%%%%%%%%%%%%%%%%%%%%%%%%%%%%%%%%%%%%%%%%%%%%%%%%%%%%%%%%%%%%%%%%%%%%%%%%%%%%
\section{Self-Consistent Dipole Model} \label{sec:SCDM}
%%%%%%%%%%%%%%%%%%%%%%%%
\begin{figure*}[!ht]
    \includegraphics{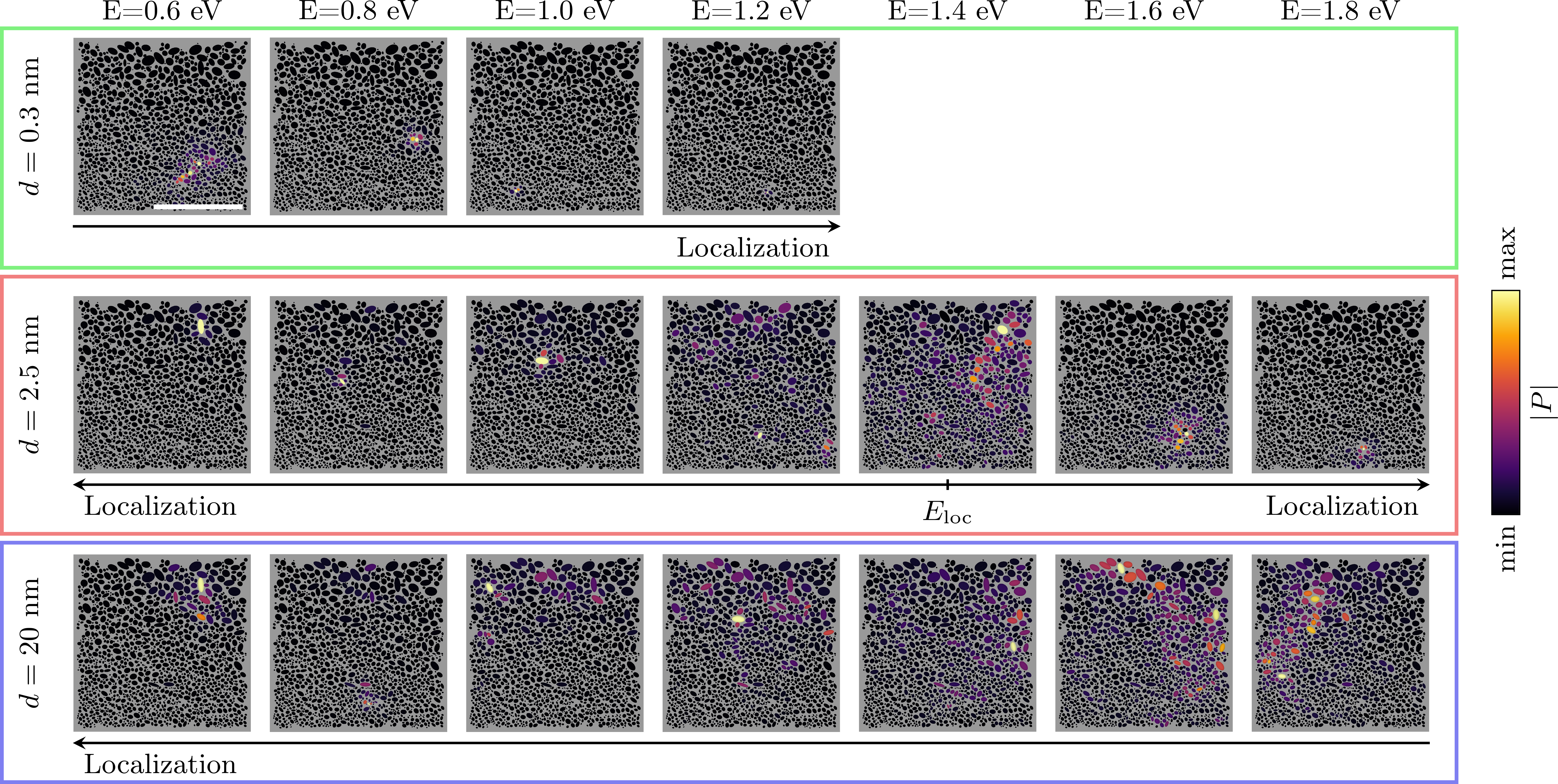}
    \caption{Simulated Polarization $|P|$ maps of selected eigenmodes with different energy and sample thickness. For thin samples ($d=0.3$\,nm, upper row/green box), localization increases the higher the energy until no resonant modes are present for energies above 1.2\,eV. Thick samples ($d=20$\,nm, lower row/blue box), on the other hand, show decreasing localization with increasing energy. The maps of samples with intermediate thickness ($d=2.5$\,nm, lower row/red box) reveal the frequency of minimal localization. Above this frequency, they behave like the thin samples, and below, they behave like the thick ones. The scale bar shown in the map with $d=0.3$\,nm and $E=0.6$\,eV corresponds to 200\,nm.}
    \label{fig:sim_maps}
\end{figure*}
%%%%%%%%%%%%%%%%%%%%%
To describe hybridized surface plasmon modes in random assemblies of gold NPs, we consider the interaction of dipolar surface plasmons of individual NPs exposed to an external electric field (the evanescent field of the beam electrons). The neglect of higher-order multipole moments leads to a slight reduction of nearest neighbor coupling, which, however, does not modify the localization behavior of the surface plasmons considered below. The coupled system of $N$ dipoles under the influence of an external electric field corresponds to the following set of $i=\{1...N\}$ equations 
\begin{equation} 
\boldsymbol{P}_{i}\left(\omega\right)-\boldsymbol{\alpha}_{i}\left(\omega\right)\sum_{j=1,\,j\neq i}^{N}\mathbf{G}_{ij}\left(\omega\right)\boldsymbol{P}_{j}\left(\omega\right)=\boldsymbol{\alpha}_{i}\left(\omega\right)\boldsymbol{E}^{\mathrm{ext}}_{i}\left(\omega\right) \label{eq:SCDM}
\end{equation}	
which needs to be solved for the individual dipole moments $\boldsymbol{P}_{i}$ of the NPs. Here, $\boldsymbol{\alpha}_{i}$ is the polarizability tensor of the individual NP $i$, $\boldsymbol{E}^{\mathrm{ext}}$ is related to the field of the beam electrons, and $\mathbf{G}_{ij}$ the retarded dipole interaction kernel that reads 
\begin{align} 
\mathbf{G}_{ij}\left(\omega\right)&=\frac{e^{ikr_{ij}}k^{2}}{r_{ij}} \cdot  \left(\mathbf{I}_3-\boldsymbol{e}_{ij}\otimes \boldsymbol{e}_{ij}\right) \nonumber \\
&+e^{ikr_{ij}}\left(1-ikr_{ij}\right)\left(\frac{3\boldsymbol{e}_{ij}\otimes\boldsymbol{e}_{ij}-\mathbf{I}_3}{r_{ij}^{3}}\right)
\end{align}
 where $k=\omega/c$ denotes the wave number, $r_{ij}$ the distance between NP $i$ and NP $j$, $\mathbf{I}_3$ the $3\times 3$ unit matrix, and $\boldsymbol{e}_{ij}$ the unit vector pointing from NP $i$ to NP $j$. 
 
 Experimentally observable resonances of such a system of coupled dipoles occur at real frequencies $\omega$, when the coefficient matrix on the left hand side of Eq.(\ref{eq:SCDM}) multiplied by $\boldsymbol{\alpha}_{i}^{-1}$ has a determinant close to zero, i.e. $\mathrm{det}\left(\boldsymbol{\alpha}^{-1}-\mathbf{G}\right)< \delta \ll 1$. Here $\boldsymbol{\alpha}$ and $\mathbf{G}$ denote the composite matrices of all $\boldsymbol{\alpha}_i$ or $\mathbf{G}_{ij}$, e.g.
 \begin{equation}
	\boldsymbol{\alpha}=\begin{pmatrix}\boldsymbol{\alpha}_{1} & \boldsymbol{0} & \cdots & \boldsymbol{0}\\
	\boldsymbol{0} & \boldsymbol{\alpha}_{2} & \ddots & \boldsymbol{0}\\
	\vdots & \ddots & \ddots & \vdots\\
	\boldsymbol{0} & \boldsymbol{0} & \cdots & \boldsymbol{\alpha}_{n}
	\end{pmatrix}.
\end{equation}
Alternatively, resonances can be identified by solving a corresponding eigenvalue problem (see Appendix\,\ref{supp. resonance}), which is computationally less expensive. Transformation into an eigenvalue problem also allows us to consider the impact of the NP thickness on the position of the resonance frequencies in a semi-analytical manner (see Appendix\,\ref{supp. diagonal} ). Employing a Drude model (plasma frequency $\omega_p$, damping parameter $\gamma$) for the dielectric function of gold and well-known expressions for the polarizability of ellipsoid NPs, we obtain the following resonance conditions,  
\begin{equation}
    \left|\lambda_n\left(\omega\right)+3\frac{\omega\left(\omega+i\gamma\right)}{\omega_p^2}-1\right|\le \delta
\end{equation}
and
\begin{equation}
    \left|\lambda_{\perp,n}\left(\omega\right)+3\frac{\omega\left(\omega+i\gamma\right)}{\omega_p^2}\right|\le \delta
\end{equation} for the eigenmodes (eigenvalues $\lambda_n$) of coupled dipole modes on a system of round NPs of random size and spatial distribution and extremely oblate NP (thin particle limit),
respectively (see again Appendix\,\ref{supp. resonance} and \ref{supp. diagonal} for details). Considering the order of magnitude of the eigenvalues  $\mathcal{O}\left(\|\lambda_n\|\right)=k^3 \prod_m a_m \ll 1$ (with $a_{1...3}$ denoting the semi-axes of the NPs) resonances in the 2D system are therefore shifted toward small frequencies with respect to the round NP limit.
To mimic the experimental situation (see Fig.\,\ref{fig:exp}\,(d)) as close as possible to the dipole model, the following procedure was applied. The HAADF STEM image contrast from the whole quadratic region (shown in Fig.\,\ref{fig:exp}\,(b) and partly in Fig.\,\ref{fig:exp}\, (d)) has been thresholded (binarized) in the first step to distinguish between gold particles and substrate. Subsequently, a particle finding algorithm that identifies particles in the first step and fits 2D ellipses to each identified particle in the second step has been applied to the binarized image (yielding Au NP coverage of 53\% and mean semi-axis size of 37\,nm). The positions of the ellipses, as well as the size and orientation of the principal axis of the ellipses, are used as input for the coupled dipole model. The remaining free parameter, the thickness of the NPs, has been varied in the simulations to study the dependency of the resonant plasmonic modes on the thickness (e.g., with respect to the shift of resonance frequency predicted above). 

In order to quantify the localization of the resonant modes, the inverse participation number has been evaluated \cite{Markel2006}. The participation number is defined through
\begin{equation}
	p_\text{sim}(\omega)=\left\langle \sum_{i=1}^{N}\frac{1}{\left|\boldsymbol{P}_{i, n}\left(\omega\right)\right|^{2}}\right\rangle _n
\end{equation}
again averaged over all resonant eigenmodes $n$ fulfilling the resonance condition. In the experiment, it is typically not possible to identify individual resonant modes due to limited energy resolution. We slightly adapted the definition of the participation number to the inverse normalized second momentum of the spectrum image
	 \begin{equation}
	  p_{\text{exp}}\left(\omega\right)=\left(\int_{-\infty}^\infty\frac{\Gamma^2\left(\boldsymbol{r}_\perp,\omega\right)}{\left(\int\Gamma(\boldsymbol{r}_\perp, \omega)d^2r_\bot\right)^2}d^2r_\bot\right)^{-1}
	 \end{equation}
within an energy interval around $\omega$, which has its roots in the quantum mechanical definition of the participation number \cite{Markel2006}.

%%%%%%%%%%%%%%%%%%%%%%%%%%%%%%%%%%%%%%%%%%%%%%%%%%%%%%%%%%%%%%%%%%%%%%%%%%%%%%%%%%%%%%%%%%%%%%%%%%%%%%%%%%%%%%%%

\section{Results}
\subsection{Simulations}
\label{subsec:sim}
In order to investigate the influence of the thickness of the NPs, several simulations were carried out (each with a homogeneous thickness of the NPs), with the thickness being changed gradually in a range between 0.2 and 20\,nm.
\begin{figure}[!ht]
    \centering
    \includegraphics{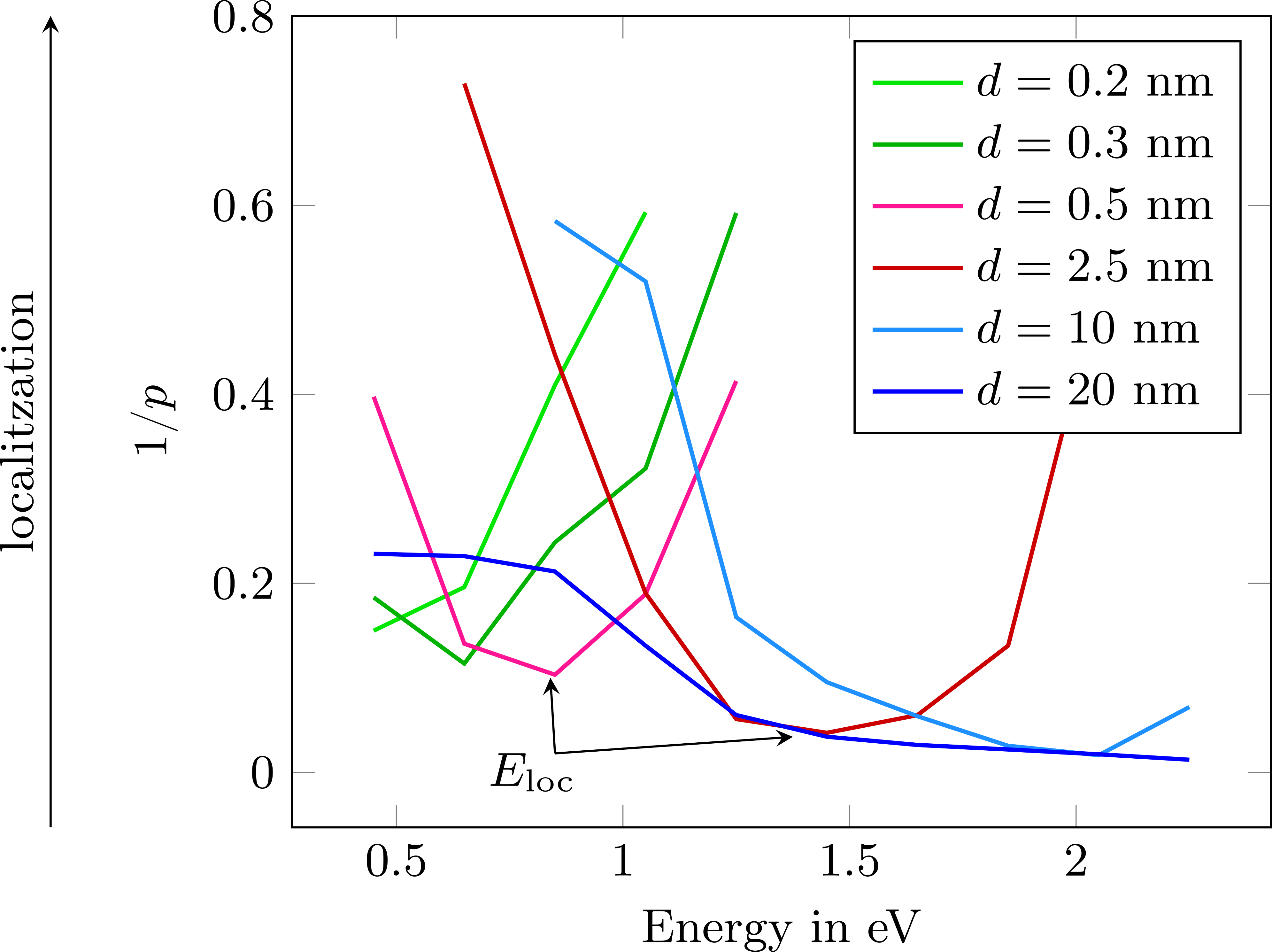}
    \caption{Inverse participation number for different thicknesses $d$ of the sample. The curves reveal a thin sample domain (green curves), a thick sample domain (blue curves) and a transition region (red curves). The minima of the curves correspond to minimal plasmon localization $E_{\text{loc}}$.}
    \label{fig:i_partn}
\end{figure}
By inspecting the spectral dependence of their localization behavior, revealed by the polarization maps (see Fig.\,\ref{fig:sim_maps}) and quantified by the inverse participation number (see Fig.\,\ref{fig:i_partn}), we can divide this thickness domain into the following subdomains:\\
(\RomanNumeralCaps{1}) a thin sample domain for NP thickness below 0.4\,nm,\\
(\RomanNumeralCaps{2}) a transition region between 0.4 and 10\,nm thickness of the NPs, and\\
(\RomanNumeralCaps{3}) a thick sample domain for NP thickness above 10\,nm.\\
In the low thickness regime (\RomanNumeralCaps{1}), the localization increases with increasing energy (see green box in Fig.\,\ref{fig:sim_maps}). Even though the dielectric function of gold allows for plasmon excitation up to $\approx2.3$\,eV, the systems show resonant modes only in a narrow frequency band (see plotted spectral range of the green curves in Fig.\,\ref{fig:i_partn}) explaining their high transparency at optical frequencies found in our previous investigations on gold networks \cite{Schultz2024Anderson}. In contrast, thick samples (regime (\RomanNumeralCaps{3})) show a decrease in localization as a function of thickness that can be observed in the polarization maps (blue box in Fig.\,\ref{fig:sim_maps}). This is reflected by the slope of the inverse participation number (blue curves in Fig.\,\ref{fig:i_partn}). Unlike the thin samples, resonant modes are present over the whole spectral range of possible plasmon excitation. Studying samples of intermediate thickness (regime (\RomanNumeralCaps{2}), red box in Fig.\,\ref{fig:sim_maps}) disclose behavior similar to thick samples up to an energy of minimal localization $E_{\text{loc}}$ of minimal localization, i.e., a minimum of the inverse participation number (see red curves in Fig.\,\ref{fig:i_partn}). Above the energy of minimal localization, the spectral localization increases again as observed in thin samples.
\subsection{Thickness Dependence of $E_{\text{loc}}$}
\begin{figure}[!ht]
    \centering
    \includegraphics{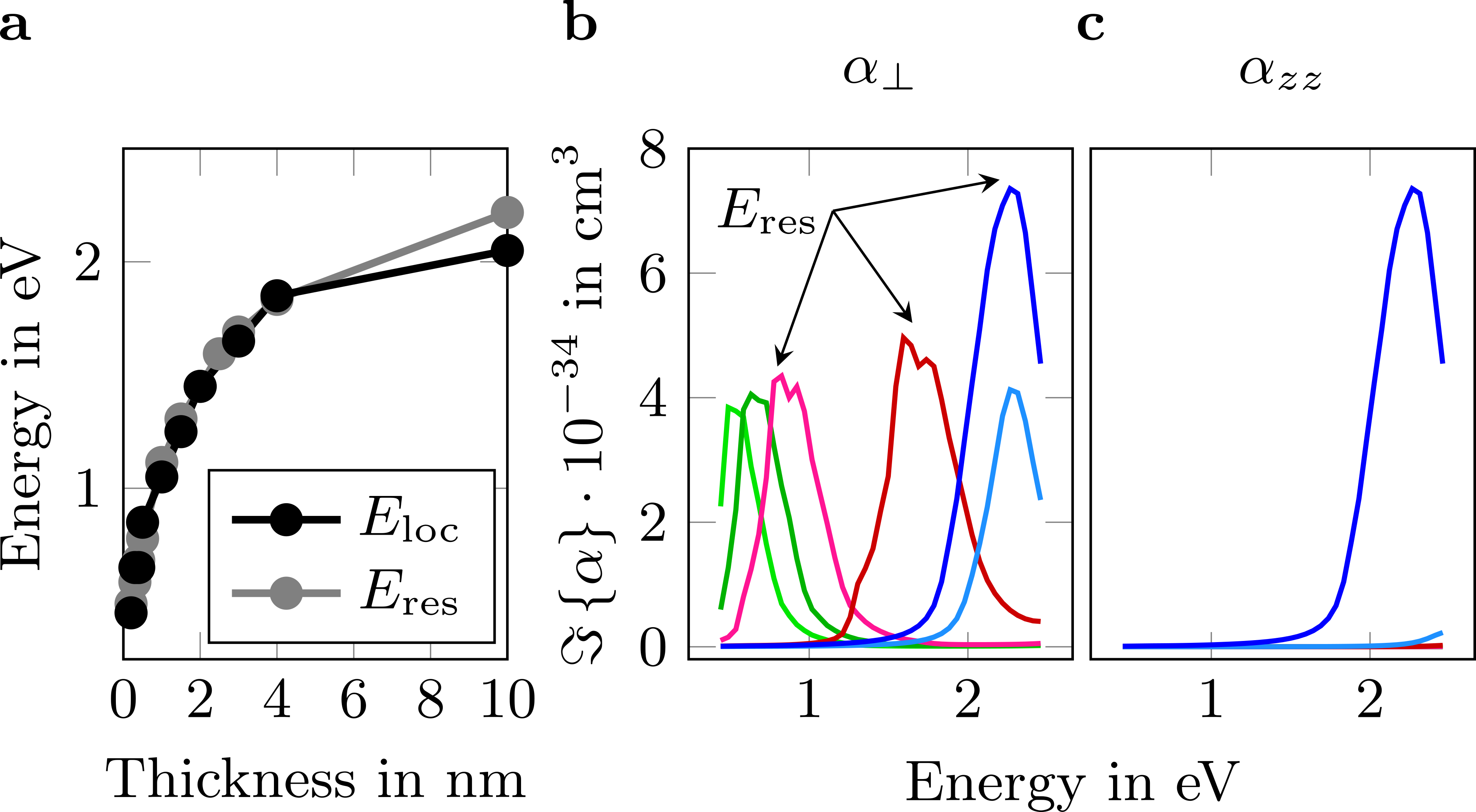}
    \caption{(a) energy of minimal localization $E_{\text{loc}}$ compared to the mean resonance energy $E_{\text{res}}$ of the dipolar mode given by the maxima of the curves in (b). (b) Mean in-plane and (c) out-of-plane imaginary part of the polarizability for different thicknesses (light green=0.2\,nm, dark green=0.3\,nm, pink=0.5\,nm, dark red=2.5\,nm, light blue=10\,nm and dark blue=20\,nm).}
    \label{fig:pol}
\end{figure}

The energy of minimal localization of the NPs increases with increasing thicknesses (see black curve in Fig.\,\ref{fig:pol} (a)). The curve ends in a plateau slightly above 2\,eV for thicknesses up to $\approx$\,10\,nm which corresponds to the transition to the thick particle limit.\\
Since the positional arrangement of the NPs was kept constant in the simulations shown in Figs.\,\ref{fig:sim_maps} and \ref{fig:i_partn}, the reason for the thickness-dependent change of the localization behavior must be the polarizability. To analyze the latter as a function of NPs thickness, we compared the mean resonance energy $E_{\text{res}}$ of the in-plane dipolar plasmon mode given by the maximum (pole) of the imaginary (real) part of the polarizability ($\alpha_\perp$) (see Fig.\,\ref{fig:pol} (b)) with the energy of minimal localization (see Fig.\,\ref{fig:pol} (a)). Here, good agreement between $E_{\text{loc}}$ and $E_{\text{res}}$ has been found for the transition region between 0.4\,nm and 10\,nm thickness. For thicknesses above 10\,nm the resonance frequency saturates similar to $E_{\text{loc}}$ around 2.3\,eV which corresponds to the upper plasmon excitation limit given by the dielectric function of the material. Accordingly, all resonant modes of the system have energies $\leq E_{\text{loc}}$ and decreasing localization with increasing energy is generally observed for these samples. In contrast, in systems with thicknesses below 0.4\,nm, all resonant modes have energies $\geq E_{\text{loc}}$ since, for lower energies, effective plasmon excitation is suppressed by dielectric damping of the material. Consequently, they show increasing localization with increasing energy.
%%%%%%%%%%%%%%%%%%%%%%%%%%%%%%%%%%%%%%%
\subsection{Experiment}
\begin{figure}[!ht]
    \centering
    \includegraphics{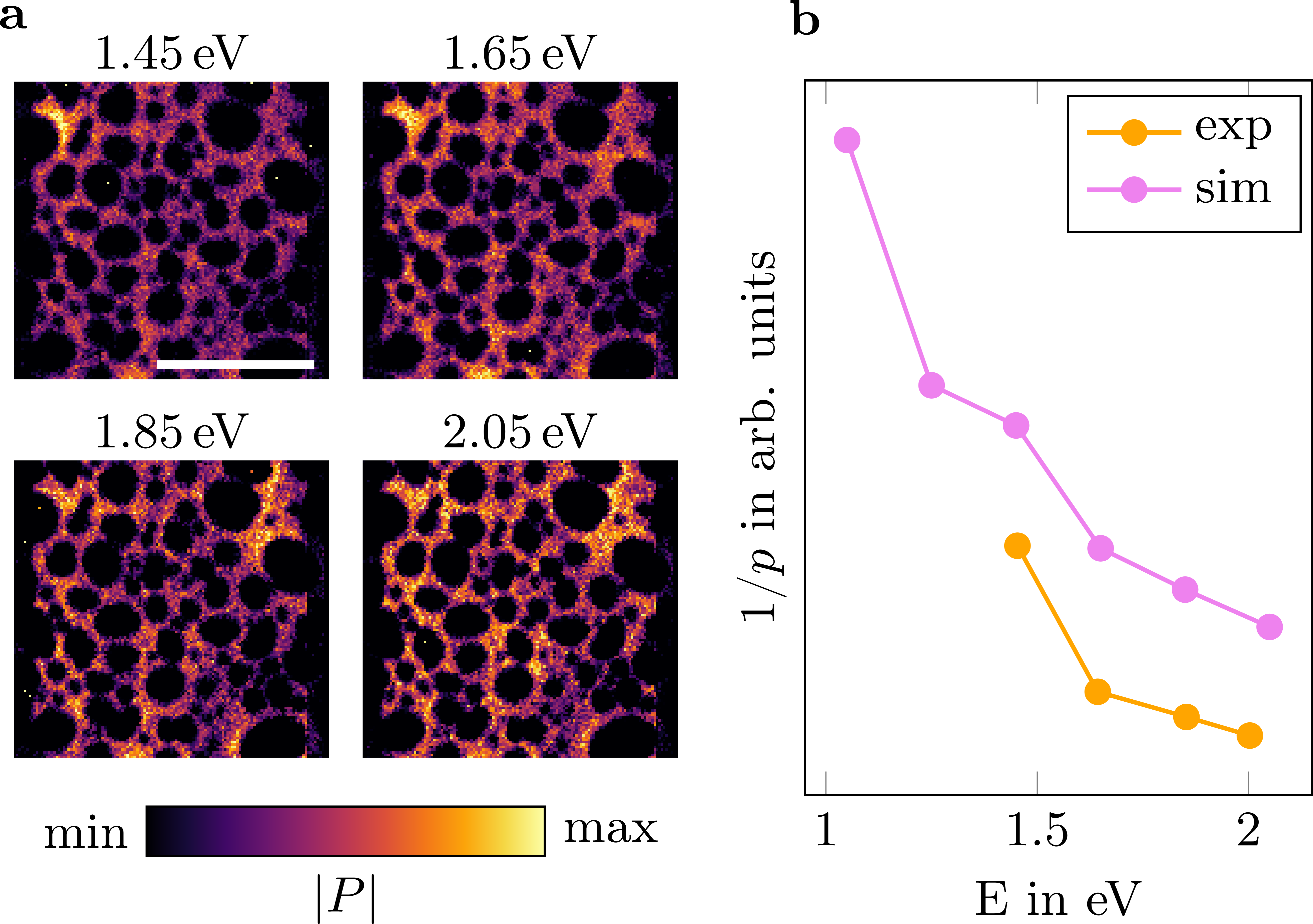}
    \caption{(a) Experimental loss probability maps for different energies reveal decreasing localization with increasing energy. This behavior is reflected by the frequency dependence of the inverse participation number (orange curve in (b)). A comparison with a simulation resembling the real scenario of random NP thickness shows good qualitative agreement. The scale bar in (a) corresponds to 200\,nm.}
    \label{fig:exp_data}
\end{figure}
In order to verify the theoretic predictions concerning the thick particle limit (the thin particle limit has been studied already in, e.g., Ref. \cite{Schultz2024Anderson}), we conducted STEM-EEL spectrum imaging on the NP assembly region indicated in Fig. \,\ref{fig:exp}. This region contains particles of average size of $\approx$ 37 nm. As the thickness of the synthesized NPs is correlated with their in-plane dimensions, the studied region consists of NPs with random thickness in the range of some 10\,nm and falls thus clearly in the thick particle limit. Accordingly, we observe decreasing plasmon localization for increasing energy in both the loss probability maps (Fig.\,\ref{fig:exp_data} (a)) and the inverse participation number (orange curve in Fig.\,\ref{fig:exp_data} (b)), as predicted with our simulation results (Sec.\,\ref{subsec:sim}), in which a constant thickness for all NPs was assumed. To model a more realistic scenario, \textit{i.e.}, random size/thickness of the NPs, we incorporate the NPs' thicknesses as the mean of their in-plane semi-axes lengths. The result of this simulation is plotted in Fig.\,\ref{fig:exp_data} (b) (violet curve) and fits remarkably well to the experimental (orange) curve.

\vspace{0.5cm}
\section{Conclusions}
We can summarize that random assemblies of plasmonic Au NPs (having a  mean semi-axis size of 37\,nm and a coverage of 53\%) exhibit a disorder-driven localization behavior that crucially depends on the thickness of the NPs. Maximal delocalization occurs at energies pertaining to a resonance condition of NPs' average dipole LSP mode (dipole mode pertaining to average LSP size) because a maximal number of NPs can contribute to a delocalized mode at such energies. This stipulates two localization regimes: one, random arrangements of NPs with thicknesses below 0.4\,nm (i.e., nanoplatelets) exhibit increasing localization over the whole plasmonically active energy range between $\approx0.3$ and 2.3\,eV, and the other one, random arrangements of NPs with thicknesses above 10\,nm show increasing delocalization in that energy regime. Intermediate thicknesses exhibit a maximal delocalization within the energy interval. These findings also translate to random metallic networks via the Babinets principle. It is also important to note that the specifics (notably energy of minimal localization) depend on the in-plane size distribution and the NP material (e.g., spectral positions change when employing Ag NPs). The observed thickness dependence of hybridized LSP modes in random 2D plasmonic NP arrangements and plasmonic thin films provides a possible tuning option for their optical response to be exploited in applications. For instance, the robust localization towards optical frequencies in the thin limit can be exploited to deliberately create highly transparent assemblies/thin films in the optical regime.

\begin{acknowledgments}
    The authors acknowledge funding from the Leibniz association within the Leibniz competition funding program under project number K524/2023. M.F.K. acknowledges the scholarship from the DAAD KOSPIE India program funded by the Federal Ministry for Economic Cooperation and Development, Germany. K. W. acknowledges the funding provided by the IFW-BAM tandem program and from the Deutsche Forschungsgemeinschaft (DFG), CRC-1415, 417590517.
\end{acknowledgments}

%%%%%%%%%%%%%%%%%%%%%%%%%%%%%%%%%%%%%%%%%%%%%%%%%%%%%%%%%%%%
\begin{appendix}

\section{Resonances of self-consistent dipole model} \label{supp. resonance}
The polarizability of an ellipsoid in the coordinate system of its principle axes (indexed by $m$) reads \cite{bohren1983}
\begin{equation}
	\alpha_{mm}\left(\omega\right)=\prod^3_{n=1}a_{n}\frac{\varepsilon\left(\omega\right)-1}{3+3L_{m}\left(\varepsilon\left(\omega\right)-1\right)}
\end{equation}
with the purely geometrical depolarization factors
\begin{equation}
	L_{m}=\frac{\prod^3_{n=1}a_{n}}{2}\int_{0}^{\infty}\frac{1}{\left(a_{m}^{2}+q\right)\sqrt{\prod_{n=1}^{3}\left(q+a_{n}^{2}\right)}}dq
\end{equation}
satisfying $L_1+L_2+L_3=1$. Here, $a_{m}$ are the semi-axes of the ellipsoid and $\varepsilon\left(\omega\right)$
the bulk dielectric function of the metal. To include radiative corrections we also incorporated the first order correction factor according to \cite{novotny_hecht_2012}, i.e., 	
\begin{equation}
	\boldsymbol{\alpha}_{i} \to \frac{\boldsymbol{\alpha}_{i}}{1-\mathrm{i}\frac{2k^3}{3}\boldsymbol{\alpha}_{i}}\,.
\end{equation}
As a consequence of $k^3 \prod_m a_m \ll 1$, however, this correction is small and may be also neglected, which we explicitly checked in the simulations by switching it on and off. In order to compute the polarizability tensor for arbitrary in-plane orientations $\theta$ of the ellipsoids (i.e., with in-plane principle axes rotated by an angle $\theta$ with respect to the coordinate axis), the diagonal $\boldsymbol{\alpha}$ tensor is transformed by left and right multiplication with an in plane rotation matrix $\boldsymbol{\alpha}_i \to \mathbf{R}(\theta_i)^\mathrm{T}\boldsymbol{\alpha}_i\mathbf{R}(\theta_i)$.

In order to separate material and geometrical parameters facilitating a more detailed discussion of the impact of several sources of disorder in the NP assembly, such as NP shape, distance, thickness, the following transformations are useful. Defining
\begin{equation}    
	\chi\left(\omega\right)=\frac{\epsilon\left(\omega\right)+2}{\epsilon\left(\omega\right)-1}
\end{equation}
allows to explicitly write the inverse of the polarizability tensor as
\begin{align}
	\boldsymbol{\alpha}^{-1}_i\left(\omega\right)&=\frac{1}{\prod_na_{i,n}}\chi\left(\omega\right)\mathbf{I}_3 \nonumber\\
	&+\frac{1}{\prod_na_{i,n}}\sum_{n=1}^3\left(3L_{n,i}-1\right)\boldsymbol{u}_n\otimes\boldsymbol{u}_n-\mathrm{i}\frac{2k^3}{3}\mathbf{I}_3
\end{align}
which again has to be transformed by rotation matrices to treat arbitrarily rotated in-plane principal axis $\boldsymbol{\alpha}_i \to \mathbf{R}(\theta_i)\boldsymbol{\alpha}^{-1}_i\mathbf{R}(\theta_i)^\mathrm{T}$. 

Defining the following tensors (containing only geometric parameters of the NPs)
\begin{equation}
	\mathbf{K}=\begin{pmatrix}\boldsymbol{\kappa}_{1} & \boldsymbol{0} & \cdots & \boldsymbol{0}\\
		\boldsymbol{0} & \boldsymbol{\kappa}_{2} & \ddots & \boldsymbol{0}\\
		\vdots & \ddots & \ddots & \vdots\\
		\boldsymbol{0} & \boldsymbol{0} & \cdots & \boldsymbol{\kappa}_{n}
	\end{pmatrix}
\end{equation}
with
\begin{equation}
	\boldsymbol{\kappa}_i = \mathbf{R}(\theta_i)\left(\sum_{n=1}^3\left(3L_{i,n}-1\right)\boldsymbol{u}_n\otimes\boldsymbol{u}_n\right)\mathbf{R}(\theta_i)^\mathrm{T}, \label{eq:kappa}
\end{equation}
\begin{equation}
	\mathbf{D}=\begin{pmatrix}\prod_{n=1}^3a_{1,n}\boldsymbol{I}_{3} & \boldsymbol{0} & \cdots & \boldsymbol{0}\\
		\boldsymbol{0} & \prod_{n=1}^3a_{2,n}\boldsymbol{I}_{3} & \ddots & \boldsymbol{0}\\
		\vdots & \ddots & \ddots & \vdots\\
		\boldsymbol{0} & \boldsymbol{0} & \cdots & \prod_{n=1}^3a_{1,N} \boldsymbol{I}_{3}
	\end{pmatrix}\,,
\end{equation}
and
\begin{equation}
	\mathbf{G}(\omega)=\begin{pmatrix}\boldsymbol{0} & \mathbf{G}_{12}(\omega) & \cdots & \mathbf{G}_{1N}(\omega)\\
		\mathbf{G}_{21}(\omega) & \boldsymbol{0} & \ddots & \mathbf{G}_{2N}(\omega)\\
		\vdots & \ddots & \ddots & \vdots\\
		\mathbf{G}_{N1}(\omega) & \mathbf{G}_{N2}(\omega) & \cdots & \boldsymbol{0}
	\end{pmatrix}\,,
\end{equation}
allows rewriting Eq. (1) of the main manuscript to
\begin{equation}
	\left( \mathbf{\chi} \left(\omega\right) -\mathbf{W}\left(\omega\right) \right) \boldsymbol{P}\left(\omega\right) = \mathbf{D}\boldsymbol{E}_{\mathrm{ext}}\left(\omega\right) \label{eq:SCDM2}
\end{equation}
with
\begin{equation}
	\mathbf{W}\left(\omega\right) = \mathbf{D}\left(\mathbf{G}\left(\omega\right)+\mathrm{i}\frac{2k^3}{3}\mathbf{I}_{3N}\right)-\mathbf{K}\,.    
\end{equation}

In this equation, terms, which only depend on the NP geometry ($\mathbf{D}$ and $\mathbf{K}$), their assembly ($\mathbf{G}$) and the material composition ($\chi$) are separated. In particular, blockdiagonal ($\mathbf{D}$ and $\mathbf{K}$) and non-diagonal ($\mathbf{G}$) sources of disorder (in the sense) can be associated to different geometric parameters.

Moreover, the equation assumes a peculiar form when making the out-plane-size of the NPs much smaller than the two in-plane dimensions (i.e., strongly oblate NPs). In that case the dipole interaction decouples into in-plane and out-of-plane (i.e. no excitation of in-plane dipole by out-of-plane and vice versa). Similarly, $\boldsymbol{\kappa}_i$ (Eq. (\ref{eq:kappa})) decomposes into an in-plane and out-of-plane part, with the in-plane one reading (taking into account $L_1=L_2=0$)
\begin{equation}
	\boldsymbol{\kappa}_{\perp i} = -\mathbf{R}_\perp(\theta_i) \mathbf{I}_2 \mathbf{R}_\perp(\theta_i)^\mathrm{T}, 
\end{equation}
and hence (from Eq. (\ref{eq:SCDM2}))
\begin{equation}
	\left( \mathbf{\chi} \left(\omega\right) - 1 - \mathbf{W}_\perp\left(\omega\right)\right) \boldsymbol{P}_\perp \left(\omega\right)= \mathbf{D}_\perp\boldsymbol{E}_{\perp,\mathrm{ext}}\left(\omega\right)  \label{eq:SCDM2D}
\end{equation}
with
\begin{equation}
	\mathbf{W}_\perp \left(\omega\right) = \mathbf{D}_\perp\left(\mathbf{G}_\perp\left(\omega\right)+\mathrm{i}\frac{2k^3}{3}\mathbf{I}_{2N}\right)\,.
\end{equation}

Consequently, the impact of $\boldsymbol{K}$ is severely simplified in this case. It particularly does not contribute to disorder anymore. We note that $\mathbf{W}$ assumes a similar shape if we assume round particles of identical size ($a_1=a_2=a_3=a$), and hence $\mathbf{K}=\mathbf{0}$ and
\begin{equation}
	\mathbf{W}_a\left(\omega\right) = \mathbf{D}\left(\mathbf{G}\left(\omega\right)+\mathrm{i}\frac{2k^3}{3}\mathbf{I}_{3N}\right)\,.   
\end{equation}

Comparing Eqs.(\ref{eq:SCDM2}) and (\ref{eq:SCDM2D}) in the constant particle size limit we note a deviation on the left-hand side, which will result in a spectral shift of the resonant modes as discussed further below.

Resonant modes of the above systems (\ref{eq:SCDM2}) and (\ref{eq:SCDM2D}) occur at (generally complex) frequencies where $ \left( \mathbf{\chi} \left(\omega\right) -\mathbf{W} \right)$ or  $ \left( \mathbf{\chi} \left(\omega\right) - 1 -\mathbf{W}_\perp \right)$ approach zero. If $\mathbf{W}$ or $\mathbf{W}_\perp$  are not defective, we can find these singularities by solving the eigenvalue problem
\begin{equation}
	\mathbf{W}\left(\omega\right)X_n\left(\omega\right)=\lambda_n\left(\omega\right) X_n\left(\omega\right)
\end{equation}
and searching for eigenvalues $\lambda_n\left(\omega\right)=\chi\left(\omega\right)$ or $\lambda_{\perp,n}\left(\omega\right)=\chi\left(\omega\right)-1$. In practice, the above condition is somewhat softened to a small resonance radius $\delta$, e.g., $\|\lambda_n\left(\omega\right)-\chi\left(\omega\right)\|\le \delta$. Moreover, we restrict the frequencies to real frequencies corresponding to real energy losses measured in the experiment. Inserting a Drude model 
\begin{equation}
	\epsilon\left(\omega\right)=1-\frac{\omega_2^2}{\omega\left(\omega+i\gamma\right)}
\end{equation}
and hence
\begin{equation}
	\chi\left(\omega\right)=1-3\frac{\omega\left(\omega+i\gamma\right)}{\omega_2^2}
\end{equation}
the resonance conditions read
\begin{equation}
	\left|\lambda_n\left(\omega\right)+3\frac{\omega\left(\omega+i\gamma\right)}{\omega_2^2}-1\right|\le \delta
\end{equation}
and
\begin{equation}
	\left|\lambda_{\perp,n}\left(\omega\right)+3\frac{\omega\left(\omega+i\gamma\right)}{\omega_2^2}\right|\le \delta
\end{equation}
respectively. Considering the order of magnitude of the eigenvalues of $k^3 \prod_m a_m \ll 1$ resonances in the 2D system are therefore shifted toward small frequencies with respect to the round NP limit.

The whole algorithm has been implemented in the Julia programming language employing efficient libraries for linear algebra and numerical integration.

\section{Diagonal versus off-diagonal disorder} \label{supp. diagonal}
\begin{figure}[!hb]
	\centering
	\includegraphics{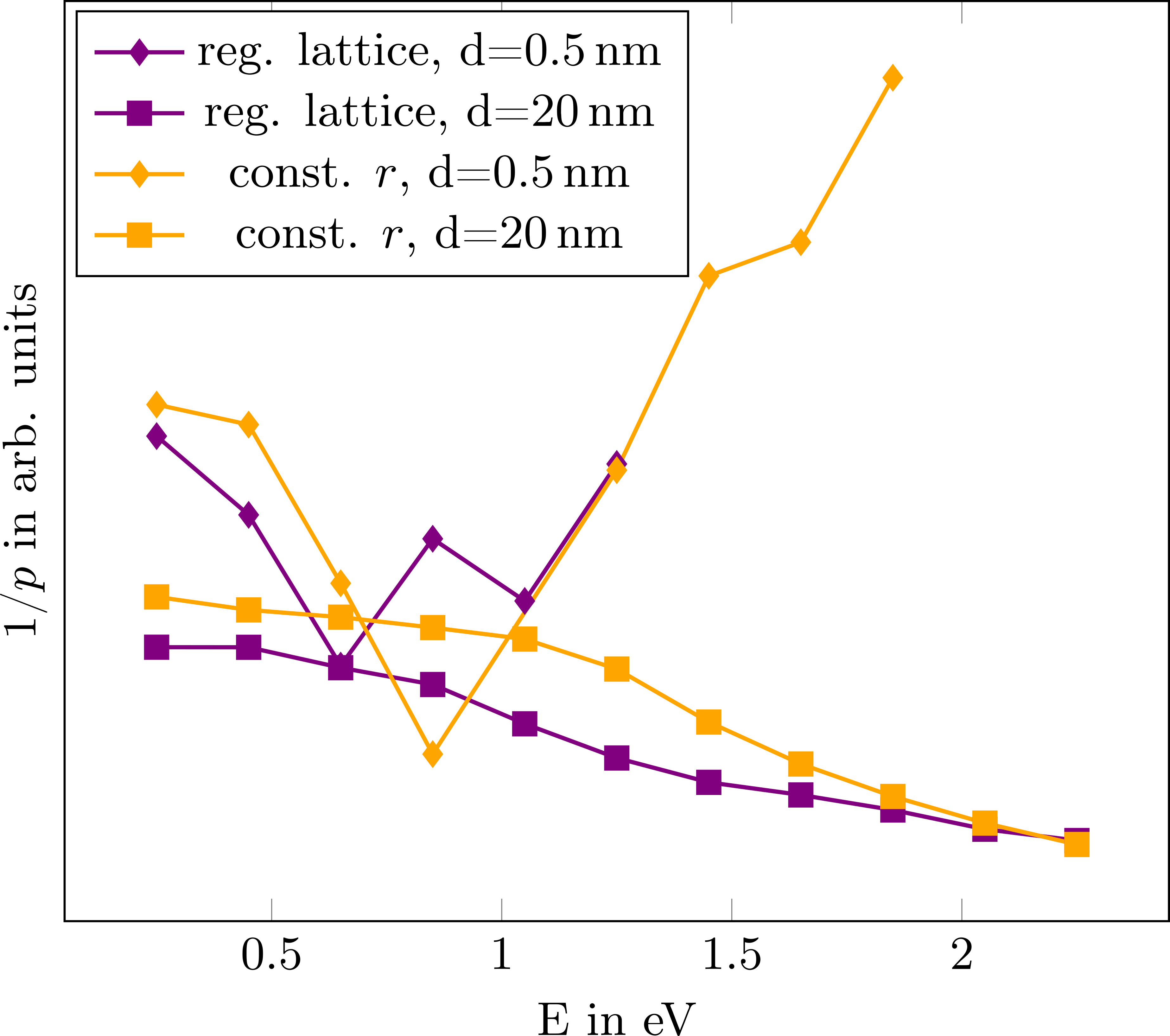}
	\caption{Inverse participation number of pure diagonal (only size randomization of NPs on a regular lattice) and off-diagonal (only position randomization of NPs of equal size) disorder in the thin (d=0.5\,nm) and thick (d=20\,nm) particle limit.}
	\label{fig:(off)diagonal disorder}
\end{figure}
Eq.\,(1) of the main manuscript and by extend Eq.\,(11) of the SI contains two types of disorder: \RomanNumeralCaps{1}) (block-)diagonal disorder originating from the size distribution of the NPs ($\mathbf{D}$ and blockdiagonal $\mathbf{K}$ in SI Eq.\,(11)), and \RomanNumeralCaps{2}) off-diagonal disorder introduced by their disordered geometrical arrangement ($\mathbf{G}$). The blockdiagional disorder $\mathbf{K}$ is suppressed in the thin particle limit as discussed previously. By individually switching off randomized particles sizes, shapes and positions, we can individually check the impact of these different disorder types in the simulations. Keeping the semi-axes length and rotation constant (no diagonal disorder)/arranging the NPs on a regular lattice (no off-diagonal disorder) while varying one of the distributions, we find that the localization behaviour is driven by both diagonal and off-diagonal disorder for the investigated geometrical parameters (i.e. sizes and distances of NPs). In other words both the random shift of resonance energies of individual NPs determined by their size (contributing to diagonal disorder) as well as their randomized coupling affects the localization (see Fig. \ref{fig:(off)diagonal disorder} of the SI).

\end{appendix}

\bibliographystyle{apsrev4-2}
\bibliography{bib}		

\end{document}